\documentclass[twocolumn,showpacs,preprintnumbers,amsmath,amssymb]{revtex4}

\usepackage{graphicx}
\usepackage{dcolumn}
\usepackage{bm}

\begin{document}


\title{ Relation between the x-dependence of Higher Twist Contribution
\\ [3mm] to $F_3$ and $g_1^p - g_1^n$
in the Light of the Recent Experimental Data }

\author{Alexander V. Sidorov}
\email{sidorov@theor.jinr.ru} \affiliation{ Bogoliubov Theoretical
Laboratory, Joint Institute for Nuclear Research, 141980 Dubna,
Russia. } %

\date{\today}

\begin{abstract} \vspace{4mm}  We compare the recent results
on the higher twist (HT) contribution to the nonsinglet
combination $g_1^p - g_1^n$ of the polarized proton and neutron
structure functions  with that one to the unpolarized structure
function $F_3$ using the assumption that the HT contributions to
the Gross-Llewellyn Smith and the Bjorken sum rules are similar.
We have found, that  the relation $\frac{1}{3}h^{F_3}(x) \approx
\frac{6}{g_A}h^{g_1^p - g_1^n}(x)$ is valid for $x \geq 0.2$ in
the case of NLO QCD approximation for the leading term parts of
the structure functions.
\end{abstract}

\maketitle


Presently the structure functions in deep inelastic lepton nucleon
scattering are  a subject of intensive experimental and
theoretical investigations. While the leading twist (LT) part of
the structure functions related with the parton distributions and
their $Q^2$-evolution is studied in detail in pQCD, the higher
twist corrections ($\sim1/Q^2$) are of a big interest and an
intensive study in the last years. From the very beginning the
x-dependence of the HT contribution was determined from analyses
of the data on the unpolarized structure functions  $F_2$
\cite{HTF2,YndSant_F2F3,AlekhinF2FL}, $F_3$
\cite{HTF3,YndSant_F2F3}, $F_1$ \cite{AleKylF1}  and $F_L$
\cite{HTFL,AlekhinF2FL}.

The most of the precise experimental data on polarized structure
functions (JLAB, HERMES, SLAC) are in the region of $Q^2 \sim 1\;
{\rm GeV}^2$. While in the determination of the parton densities
(PD) in the unpolarized case we can cut the low $Q^2$ and $W^2$
data in order to eliminate the less known non- perturbative HT
effects, it is impossible to perform such a procedure for the
present data on the spin- dependent structure functions without
loosing too much information. That is why the higher twist effects
are especially important in the case of polarized structure
functions and should be taken into account in the QCD analysis.
Since the first results on HT contribution to the $g_1$ structure
function \cite{LSS_HT} the accuracy of the x-dependence of the HT
in $g_1^p$ and $g_1^n$ extracted directly from the data has been
considerably improved \cite{LSS07}.

Are there any relations between the HT contributions to the
different structure functions? We continue to discuss this
question based on the new results of the paper \cite{LSS07} on the
x-dependence of the HT in $g_1$. It was obtained using the recent
very precise CLAS \cite{CLAS06} and COMPASS \cite{COMPASS06}
inclusive polarized DIS data. In this note we continue our study
\cite{GLSvsBj} on the relation between the HT contribution to the
unpolarized structure function $F_3$ and $g^p_1-g^n_1$ which are
pure non-singlets.

 As it was shown in the paper \cite{Kodaira} the $Q^2$-evolutions of
the $F_3$ and the nonsinglet part of the  $g_1$ structure
functions are identical up to NLO. Moreover, the x-shapes of the
$F_3$ and nonsinglet part of $g_1$ are also similar. By analogy,
one could suppose that the HT contributions to $F_3$ and
$g^p_1-g^n_1$ are similar too. Such an assumption was used for the
first moments of the HT corrections in the Gross-Llewellyn Smith
and Bjorken sum rules in the infrared renormalons approach
\cite{KataevGLSBjp}:
\begin{widetext}
\begin{eqnarray}
GLS(Q^2) &=&  \int^1_0 dx F_3 (x,Q^2)
 =  3(GLS_{pQCD}-\frac{\langle\langle O_1 \rangle\rangle}{Q^2})   \label{GLS}  \\
Bjp(Q^2) &=& \int^1_0 dx [g^{p}_1 (x,Q^2)-g^{n}_1 (x,Q^2)]
 = \frac{g_A}{6}(Bjp_{pQCD}-\frac{\langle\langle O_2
\rangle\rangle}{Q^2})   \label{Bjp}
\end{eqnarray}
\end{widetext}
where
\begin{eqnarray}
\langle\langle O_1 \rangle\rangle & \approx &\langle\langle O_2
\rangle\rangle
 \label{HTGLSBJ}
\end{eqnarray}

Here $GLS_{pQCD}$ and $Bjp_{pQCD}$ are the leading twist
contribution to corresponding sum rules:
\begin{eqnarray}
GLS_{LO}=Bjp_{LO}& = & 1 \\
GLS_{NLO}=Bjp_{NLO} & = & 1-\alpha_S(Q^2)/\pi
 \label{SRlonlo}
\end{eqnarray}

\begin{figure}[t]
\includegraphics[scale=0.75]{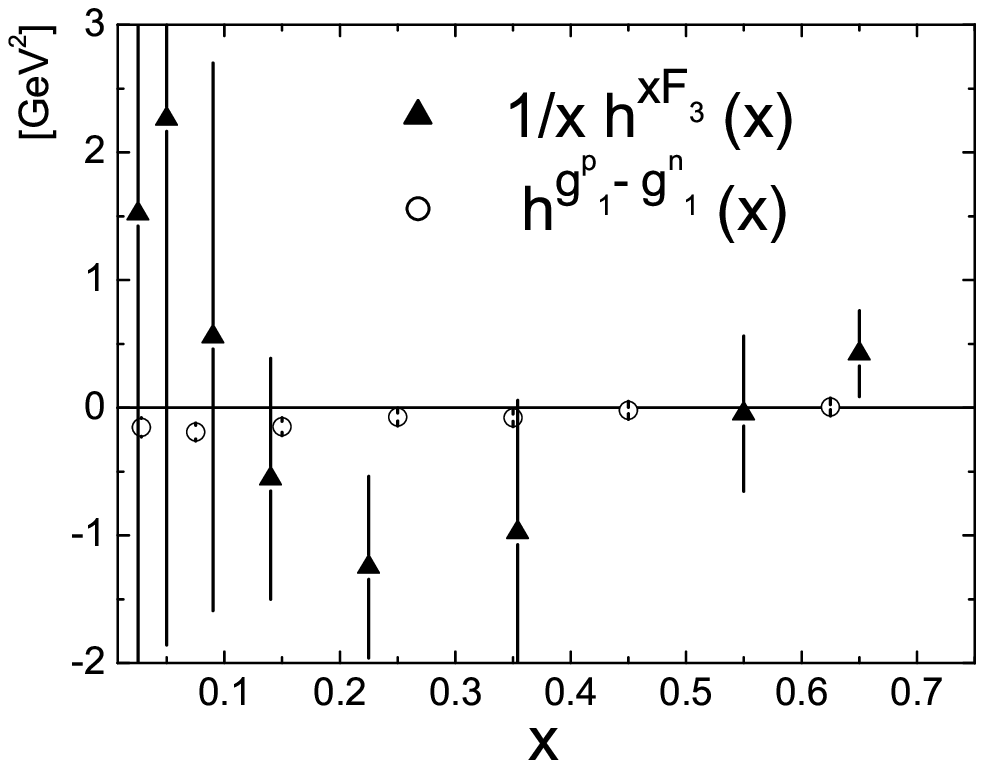}
 \caption{Comparison of the  NLO($\rm \overline{MS}$)
 results for $h^{F_3}(x)$
based on the analysis of the CCFR data \cite{KPS,CCFRdata} -
(triangles), and for $h^{g_1^p - g_1^n}(x)$ based on the results
of \cite{LSS07} - open cycles.
 \label{htg1f3}}
\end{figure}

\begin{figure}[t]
\includegraphics[scale=0.75]{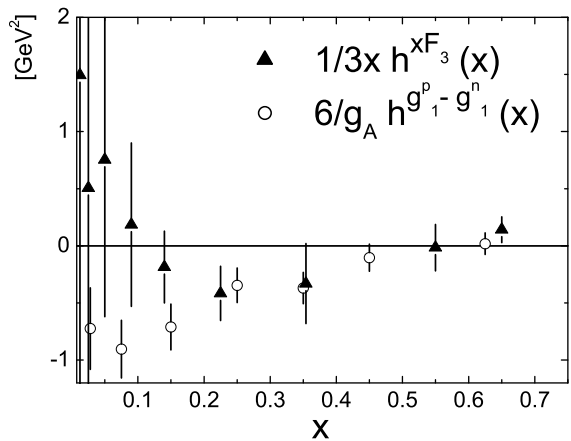}
 \caption{Comparison of the  NLO($\rm \overline{MS}$)
 results for $\frac{1}{3x}h^{xF_3}(x)$
based on the analysis of the CCFR data \cite{KPS,CCFRdata} -
(triangles), and for $\frac{6}{g_A}h^{g_1^p - g_1^n}(x)$ based on
the results of \cite{LSS07} - open cycles.
 \label{htglsBj}}
\end{figure}

In this note we are going to verify if the relation (3) between
the lowest moments of the HT contribution can be generalized for
the higher twists themselves, namely:
\begin{eqnarray}
\frac{1}{3x}h^{xF_3}(x)& \approx & \frac{6}{g_A}h^{g_1^p -
g_1^n}(x)
 \label{HTF3g1}
\end{eqnarray}

To test this relation we will use the values of HT obtained in the
QCD analysis of the corresponding structure functions in model
independent way. In the QCD analysis of DIS data when the higher
twist corrections are taken into account, the structure functions
are given by:
\begin{eqnarray}
xF_3 (x,Q^2) & = & xF_3(x, Q^2)_{\rm LT}
+ h^{xF_3}(x)/Q^2~  \nonumber \\
g^{p(n)}_1(x, Q^2) &  = & g^{p(n)}_1(x, Q^2)_{\rm LT}  +
h^{g_1^p(g_1^n)}(x)/Q^2 \label{LTHT}
\end{eqnarray}

In (\ref{LTHT}) $h^{xF_3}(x)$, $h^{g_1^p}(x)$ and $h^{g_1^n}(x) $
are the {\it dynamical} higher twists corrections to $xF_3$,
$g^p_1$ and $g_1^n$, which are related to multi-parton
correlations in the nucleon. They are non-perturbative effects and
can not be calculated without using models. The target mass
corrections, which are also corrections of inverse powers of
$Q^2$, are calculable {\cite{WW,TB} and effectively belong to the
leading twist term. A model independent determination of
$h^{xF_3}(x)$ was done in \cite{KPS} on the basis of the analysis
of CCFR-NuTev (anti-)neutrino deep--inelastic scattering data
\cite{CCFRdata} at $Q^2 \geq 5~GeV^2$. The values of
$h^{g_1^p}(x)$ and $h^{g_1^n}(x)$ in  NLO($\rm \overline{MS}$) are
given in \cite{LSS07}, where the results of the analysis of the
world data on polarized structure function $g_1$  at $Q^2 \geq
1~GeV^2$,  are presented including the precise CLAS \cite{CLAS06}
and COMPASS \cite{COMPASS06} $g_1/F_1$ data. This analysis
provides more precise and detailed results on $h^{g_1^p}(x)$ and
$h^{g_1^n}(x)$. In particular, the x-range is split into 7 bins
instead of 5, as used in the previous analyses \cite{LSS_HT,JHEP}.
Using these new results and taking into account the coefficients
in (\ref{GLS}) and (\ref{Bjp}) one could construct the l.h.s. and
r.h.s. of Eq. (\ref{HTF3g1}).

In Fig. \ref{htg1f3} we present the HT contributions to $F_3$
structure functions and to the nonsinglet combination $g_1^p -
g_1^n$. One can see the large difference of the scales for HT
contribution to polarized (small central values) and unpolarized
(large central values) structure functions.

In Fig. \ref{htglsBj}  the coefficients in (\ref{GLS}) and
(\ref{Bjp}) are taking into account.  As seen from Fig.
\ref{htglsBj}, the equality (\ref{HTF3g1}) is approximately valid
for $x \geq 0.2$. It means that the higher Mellin moments of the
both parts of equation (\ref{HTF3g1}) should approximately
coincide:
\begin{eqnarray}
\int^1_0 dx ~x^N \frac{1}{3x}h^{xF_3}(x)&  \approx & \int^1_0 dx
~x^N \frac{6}{g_A}h^{g_1^p - g_1^n}(x),  \nonumber  \\ ~N - large.
 \label{momHTF3g1}
\end{eqnarray}

In oder to show the similarity of the functions
$\frac{1}{3x}h^{xF_3}(x)$ and $\frac{6}{g_A}h^{g_1^p - g_1^n}(x)$,
we have parametrised them in the region $x>0.2$ by linear
function: $A+Bx$. The results of parametrisation are in a good
agreement for the both constants, A and B (\ref{lininterF3},
\ref{lininterg1}).

\begin{widetext}
\begin{eqnarray}
\frac{1}{3x}h^{xF_3}(x) &  = &[(-0.75 \pm 0.33) +(1.36 \pm
 0.58)x]~GeV^2,   \label{lininterF3} \\
 \frac{6}{g_A}h^{g_1^p - g_1^n}(x)& = &[(-0.66 \pm 0.21) +(1.10
\pm
 0.42)x]~GeV^2, \label{lininterg1} \\
  & & x>0.2
 \nonumber
\end{eqnarray}
\end{widetext}

We would like to mention, that equality (3) is suggested in the
framework of the infrared renormalon approach, so the violation of
equality (4), which is shown in Fig. \ref{htglsBj} at $x < 0.2$,
could be due to the contribution of the dynamical higher twists
connected with the non-perturbative structure of the nucleon in
this $x$ region.

Finally, we would like to note, that there are additional sources
of uncertainties which should be taken into account in a more
detailed test of Eq. (\ref{HTF3g1}): the contribution of ${\cal
O}(1/Q^4)$); the separation of the twist-3 contribution in the
polarized case, which is effectively included in $h^{g_1}(x)$; the
$Q^2$ dependence of the functions $h(x)$, etc. \\ [4mm]

\begin{acknowledgments}
This research was supported by the RFBR grants N~05-02-17748,
N~07-02-01011 and N~07-02-01046.
\end{acknowledgments}

\end{document}